\documentclass[aps,pre, twocolumn,showpacs, amsmath,amssymb, superscriptaddress, reprint, longbibliography,floatfix]{revtex4-1}
\usepackage[utf8]{inputenc}
\usepackage[T1]{fontenc}
\usepackage{xcolor,amsmath,amssymb}
\usepackage{graphicx}
\usepackage{enumerate}
\usepackage{wasysym}
\usepackage[textsize=footnotesize]{todonotes}

\definecolor{darkblue}{rgb}{0,0,0.6}
\definecolor{darkred}{rgb}{0.6,0,0}
\usepackage[colorlinks=true,urlcolor=darkblue, citecolor=darkblue, linkcolor=darkred, hyperfootnotes=false]{hyperref}

\newcommand{\ee}{\boldsymbol{e}}

\newcommand{\nn}{\boldsymbol{n}}

\newcommand{\rr}{\boldsymbol{r}}

\newcommand{\vv}{\boldsymbol{v}}

\newcommand{\transp}{^\mathrm{T}}

\begin{document}

\title{Dynamics of a self-propelled particle in a harmonic trap}

\author{Olivier Dauchot}
\affiliation{Gulliver, UMR CNRS 7083, ESPCI Paris, PSL Research University, 10 rue Vauquelin, 75005 Paris, France}

\author{Vincent Démery}
\affiliation{Gulliver, UMR CNRS 7083, ESPCI Paris, PSL Research University, 10 rue Vauquelin, 75005 Paris, France}
\affiliation{Univ Lyon, ENS de Lyon, Univ Claude Bernard Lyon 1, CNRS, Laboratoire de Physique, F-69342 Lyon, France}

\begin{abstract}
The dynamics of an active walker in a harmonic potential is studied experimentally, numerically and theoretically.
At odds with usual models of self-propelled particles, we identify \emph{two} dynamical states for which the particle condensates at finite distance from the trap center. In the first state, also found in other systems, the particle points radially outward the trap, while diffusing along the azimuthal direction. In the second state, the particle performs circular orbits around the center of the trap. We show that self-alignment, taking the form of a torque coupling the particle orientation and velocity, is responsible for the emergence of this second dynamical state. The transition between the two states is controlled by the persistence of the particle orientation. 
At low inertia, the transition is continuous. 
For large inertia the transition is discontinuous and a coexistence regime with intermittent dynamics develops. 
The two states survive in the over-damped limit or when the particle is confined by a curved hard wall.
\end{abstract}

\maketitle
Understanding the effect of confinement on active systems has become a central topic of research in active matter. The motivation is twofold. 
On one hand living organisms and self-propelled particles rarely evolve in an infinite and unbounded medium~\cite{Bechinger:2016cf}; understanding their behavior in confined environment is key to the understanding of realistic systems. 
On the other hand confining self-propelled particles within an external trap is an excellent mean of analyzing the dynamics and the mechanical pressure developed by such systems~\cite{Fily:2014gy,Yan:2015cf,Solon:2015hza,Nikola:2016jca,DeDier:2016he,Junot:2017fd,Deblais:2018dv}.

The mean square displacement of an isolated particle is the same for the three main models of self-propelled particles introduced so far, namely Active Brownian Particles (ABP), Run and Tumble Particles (RTP) and Active Ornstein-Uhlenbeck Particles (AOUP)~\cite{Solon:2015jd,Fodor:2018en}. 
It is therefore not a discriminating observable. 
The steady state density profile in a confining potential is a far more informative quantity, which, as such, has recently received considerable attention~\cite{Solon:2015hza,Maggi:2015jl}. 
Apart from the notable exceptions of RTPs in one dimension and AOUPs in a harmonic trap, the exact steady state is theoretically unknown. 
However, theoretical approximations together with numerical simulations~\cite{Solon:2015hza,Maggi:2015jl,Basu:2018wo} and experiments with colloids in an acoustic trap~\cite{DeDier:2016he} all show that self-propelled particles confined in a 2d axisymmetric trap accumulate at a finite distance of the trap center when the trap stiffness is large. 
The reason is that the self-propelled particle climbs the potential until it gets stuck with its orientation pointing radially outward of the trap, at a distance much shorter than the persistence length of its isolated dynamics.

\begin{figure}[t]
\vspace{2mm}
\begin{center}
\includegraphics[scale=.9]{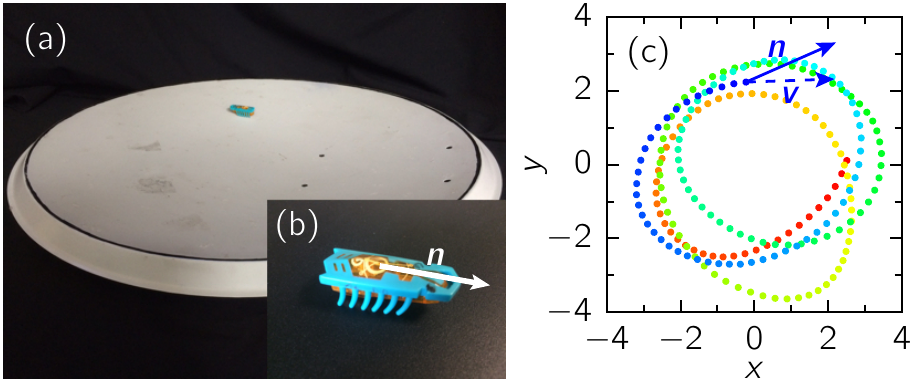}
\end{center}
\vspace{-5mm} 
\caption{HexBug-Nano (b) (https://www.hexbug.com/nano) running in a parabolic dish (a).
(c) Observed orbiting trajectory, the arrows represent the particle orientation $\nn$ and velocity $\vv$.
}
\label{fig:hexbug}
\vspace{-5mm} 
\end{figure}

This is however in sharp contrast with the simplest experiment one can think of: observing a hexbug-nano~\cite{hexbug} in a parabolic antenna. As can be seen from Fig.~\ref{fig:hexbug}, the hexbug orbits around the trap center. 
The main goal of this letter is to examine the reasons for such a discrepancy. 
Possible factors are inertia, which until recently~\cite{Scholz2018} has been overlooked in the active matter literature, and self-alignment~\cite{Weber:2013bj,LolandBore:2016if}, which has been shown to be responsible for the collective motion of isotropic walkers~\cite{NguyenThuLam2015}. 
We show that this self-alignment is the key ingredient for observing such rotating orbits. Furthermore, we demonstrate that there is a transition from the ``orbiting'' state to the ``climbing'' state depending on the persistence time of the particle orientation. Finally, the nature of the transition is controlled by inertia. 
For large relaxation time of the velocity, the transition is discontinuous and an intermittent dynamics between the two states is observed. For small relaxation time, the transition is continuous. Remarkably the transition subsists in the over-damped limit.

The experimental set up is composed of one hexbug, a toy robot~\cite{hexbug}, confined within a parabolic dish (Fig.~\ref{fig:hexbug}). 
The parabola (Grundig QGP 2400) has an elliptic section with a small axis of $580$~mm, a long axis of $630$~mm and is $50$~mm deep.
The hexbug motion is tracked at $25$~Hz using a standard CCD camera.  
The hexbug has a length $L=45$~mm, a characteristic width $15$~mm and height $15$~mm and a mass $m=7.5$~g, including the battery. 
Its motion is obtained from an internal vibration (frequency $f_0=115$~Hz and acceleration $\Gamma_0 \simeq 1.5g$) transmitted to twelve legs, the shape of which ensures propulsion at a typical speed $v_0=100$ mm$/$s. 
The persistence length of the motion exceeds one meter. 
An alternative way to obtain motion for the hexbug is to shake it vertically at the same frequency $f_0$, while keeping the engine switched off. 
The parabola is thus attached to a vibrating shaker, which provides a sinusoidal motion with an acceleration $\Gamma \in [1-3]g$. Note that the vibration amplitude not only ensures the deterministic motion of the hexbug, but also contributes to the stochastic part of it. In the following we choose $L$ as the unit length, while the time is expressed in seconds.

\begin{figure}[t]
\vspace{0mm}
\begin{center}
\includegraphics[scale=.9]{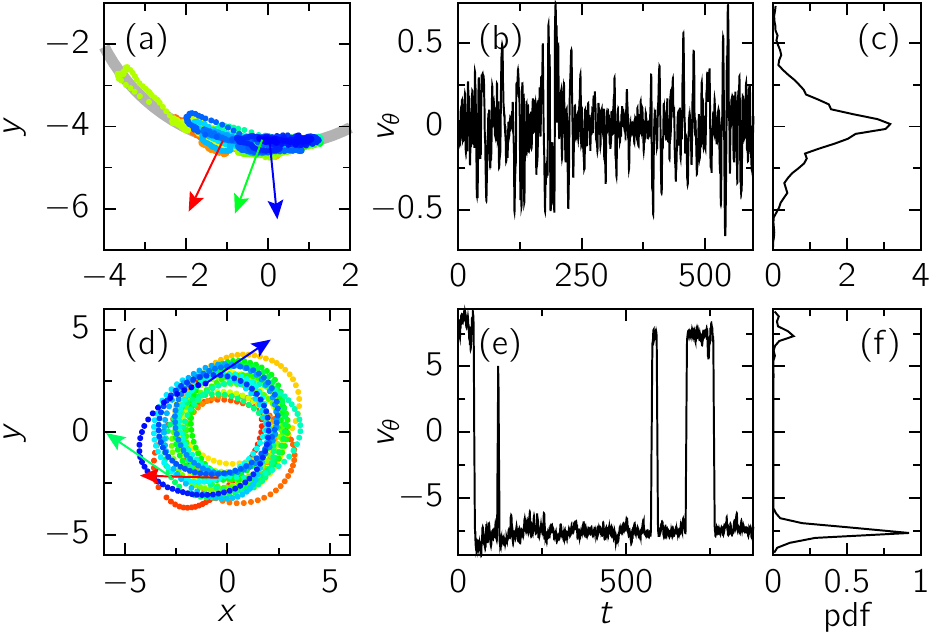}
\end{center}
\vspace{-5mm} 
\caption{{\bf Experimental dynamical regimes.}
(a-c): Climbing motion: the hexbug is stuck, facing the slope, at a given height in the potential and diffuses laterally; engine on, $\Gamma = 1.8g$.
(d-f): Orbiting motion; the hexbug rotates around the center of the parabola; engine off, $\Gamma =1.0g$.
(a,d): Trajectory; the color codes the time, the orientation of the hexbug body is shown with arrows at three different times.  
(b,e): Azimuthal velocity as a function of time. 
(c,f). Probability distribution of the azimuthal velocity.}
\label{fig:exp1}
\vspace{-5mm} 
\end{figure}

Our main experimental finding is that there is a transition from an orbiting motion  to a climbing motion when, the vibration being switch on, the propulsive power of the hexbug decreases as battery gets low. Figure~\ref{fig:exp1} illustrates the two regimes as obtained in ``steady state'' conditions : (i) engine on with a new battery and (ii) engine off with a rather low level of vibration. In the orbiting regime, the hexbug rotates around the bottom of the parabola, sometimes reverting its orbiting direction. 
The radial position fluctuates significantly. The velocity $\vv$ and the orientation of the hexbug longitudinal axis $\nn$ are closely aligned, with $\frac{\vv}{\|\vv\|} \cdot \nn > 0.8$. In the climbing regime, the hexbug orientation points radially away from the center of the parabola. The motion is localized on a well defined radial position. The azimuthal velocity fluctuates around zero and the angular motion is diffusive.
The transition is essentially independent from the vibration amplitude as long as $\Gamma > 1.5g$. 
The same transition is observed when the engine of the hexbug is switched off and the vibration amplitude $\Gamma$ is increased from $1$ to $2$g. The precise location of the transition is hard to determine, the reason being that there is a range of vibration for which there is actually a coexistence of the two dynamics, leading to complex intermittent motion.

In order to better examine this transition and identify the proper control parameters, we adopt a description of self-propelled particle inspired from the one first introduced to describe granular walkers~\cite{Weber:2013bj}:~the velocity $\vv$ and orientation $\nn$ obey
\begin{align}
m\dot\vv&=F_0\nn-\gamma\vv-\kappa \rr, \label{eq:motion_v_dim}\\
\tau\dot\nn&=\zeta(\nn\times\vv)\times\nn+\sqrt{2\alpha}\xi\nn_\perp. \label{eq:motion_n_dim}
\end{align}
Equation (\ref{eq:motion_v_dim}) contains the mass of the hexbug $m$, the self-propulsive force $F_0$, the friction coefficient $\gamma$, and the stiffness $\kappa$ of the harmonic potential; in the absence of confinement, the hexbug thus moves with a velocity $v_0=F_0/\gamma$.
The orientation dynamics (Eq.~(\ref{eq:motion_n_dim})) is over-damped and contains the key ingredient, specific to the model, namely the presence of a self-aligning torque of the orientation $\nn$ towards the velocity $\vv$. This torque originates from the fact that the dissipative force is not symmetric with respect to the propulsion direction $\nn$ when $\vv$ is not aligned with $\nn$. This ingredient was shown to be at the root of the emergence of collective motion in a system of vibrated polar discs~\cite{Weber:2013bj,Lam:2015bp}.
Finally, the orientation dynamics contains a Gaussian noise $\xi(t)$ with correlations $\langle \xi(t)\xi(t') \rangle=\delta(t-t')$; $\alpha/\tau^2$ is the rotational diffusion coefficient.

Rescaling length by $r_0=F_0/\kappa$ and time by $t_0=\gamma/\kappa$, we arrive at the dimensionless equations of motion  
\begin{align}
\tau_v\dot\vv&=\nn-\vv-\rr,\label{eq:motion_v}\\
\tau_n\dot\nn&=(\nn\times\vv)\times\nn+\sqrt{2 D}\xi\nn_\perp, \label{eq:motion_n}
\end{align}
which contain only three parameters, $\tau_v=m\kappa/\gamma^2$, $\tau_n= \tau\kappa/(\zeta F_0)$, and $D=\alpha\gamma\kappa/(\zeta F_0)^2$.
Note that $\tau_v$ does not depend on $F_0$, while $\tau_n$ increases when $F_0$ decreases.


\begin{figure*}[t!]
\begin{center}
\includegraphics[scale=.9]{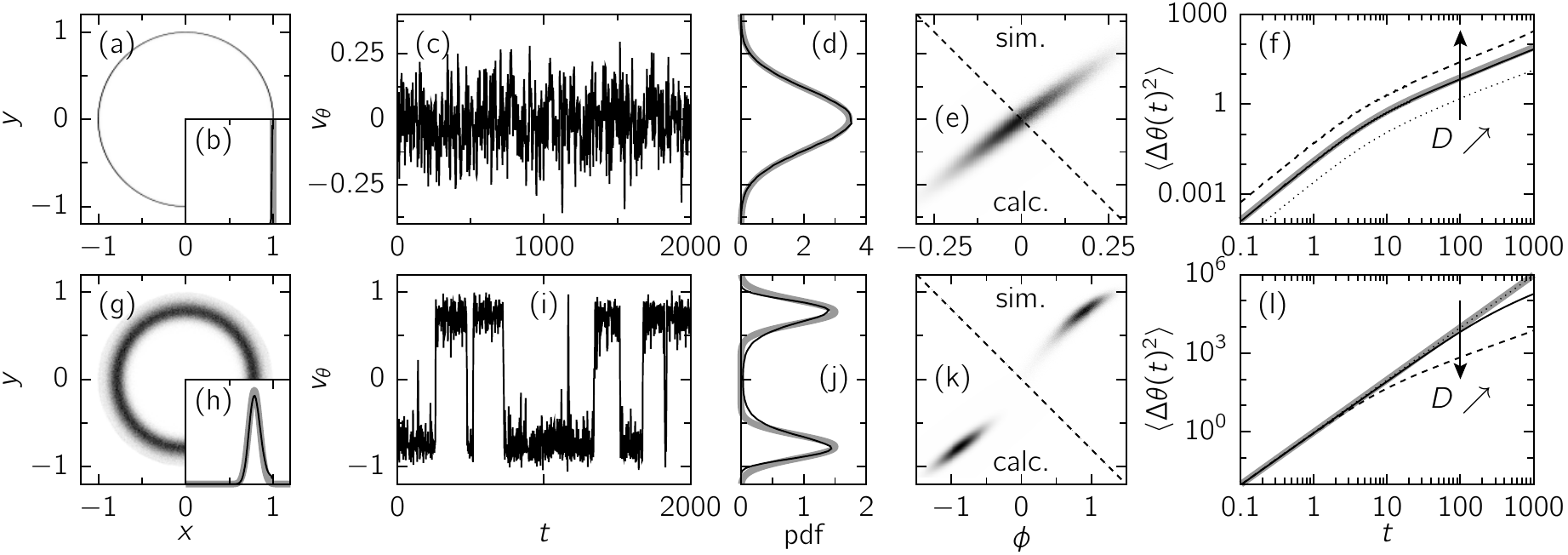}
\end{center}
\vspace{-5mm}
\caption{{\bf Numerical dynamical regimes.} Simulations for $\tau_v=0.2$, $D=0.01$ and (a-f) climbing dynamics, $\tau_n=1.5$; (g-l) orbiting dynamics, $\tau_n=0.5$.
(a,g): $(x,y)$ probability density function (pdf), (b,h): pdf of the distance to the parabola center.
(c,i): azimuthal velocity as a function of time.
(d,j): pdf of the azimuthal velocity (thin black lines) and theoretical prediction at weak noise (thick gray lines).
(e,k): pdf in the orientation-azimuthal velocity plane from simulations (top right) and weak noise calculation (bottom left).
(f,l): mean squared displacement (MSD) of the angular coordinate $\theta$ (thin solid line) and theoretical prediction (thick gray line). Dotted and dashed lines: MSD for $D=0.002$ and $D=0.05$, respectively. }
\label{fig:numerics}
\end{figure*}

We simulate Eqs.~(\ref{eq:motion_v}, \ref{eq:motion_n}) with a small noise amplitude $D=0.01$ and recover the behavior observed in the experiments: for different values of $\tau_n$ and $\tau_v$, the particle adopts either a climbing ($\tau_v=0.2$, $\tau_n=1.5$, Fig.~\ref{fig:numerics}(a-f)) or an orbiting dynamics ($\tau_v=0.2$, $\tau_n=0.5$, Fig.~\ref{fig:numerics}(g-l)).
In both cases the particle is localized at a finite distance from the center of the parabola, but the distribution is sharply peaked at $r=1$ in the climbing state, while it is more broadly distributed in the orbiting state (Fig.~\ref{fig:numerics}(b,h)).
In the climbing state, the azimuthal velocity fluctuates around 0, while, in the orbiting state, it fluctuates around two opposite finite values, the sign changes corresponding to spontaneous inversions of the rotation direction (Fig.~\ref{fig:numerics}(c,d,i,j)). 
In both cases, the direction of motion is directly correlated to the orientation of the particle expressed in polar coordinates $\nn=\cos(\phi)\ee_r+\sin(\phi)\ee_\theta$ (Fig.~\ref{fig:numerics}(e,k)). 
In the climbing state $\nn$ fluctuates around the radial orientation $\ee_r$. 
In the orbiting case, there is a finite angle between $\nn$ and $\vv$ : $\vv \parallel \ee_\theta$ and $\nn$ is always pointing outwards because of the finite relaxation time $\tau_n$. 
Finally, one can characterize the angular dynamics with the mean square angular displacement (MSD) $\langle \Delta\theta(t)^2 \rangle$ with $\Delta\theta(t)=\theta(t)-\theta(0)$ (Fig.~\ref{fig:numerics}(f,l)). 
Not surprisingly, in both cases, the MSD is ballistic at short time and crosses over to a diffusive behavior at long time. 
More interesting is the dependence on the angular noise: for the climbing dynamics, the noise amplitude affects the magnitude of the velocity and thus of the MSD, leaving the crossover time unchanged; for the orbiting dynamics, increasing the noise enhances the rate of inversion of the rotation direction and thus reduces the crossover time, thereby reducing the long-time diffusion coefficient.

\begin{figure}.
\vspace{-5mm}
\begin{center}
\includegraphics[scale=.9]{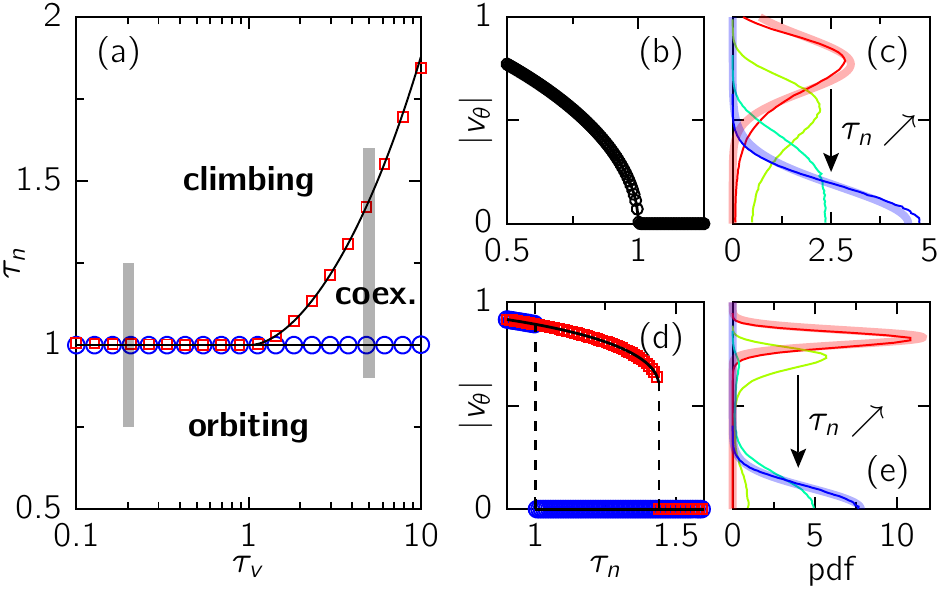}
\end{center}
\caption{{\bf Noiseless steady states.} (a) Phase diagram; (the thick gray lines indicate the values of the parameters $(\tau_v,\tau_n)$ used in the right panels.  
(b, d) Azimuthal velocity as a function of $\tau_n$ for $\tau_v=0.2$ (b) and $\tau_v=5$ (d), where the symbols represent the steady state obtained from two different initial conditions $IC_1$ (blue circles) and $IC_2$ (red squares) as defined in the text.
(c, e) Distribution of azimuthal velocity through the orbiting (red) to climbing (blue) transition for $D=0.01$ for $\tau_v=0.2$ and $\tau_n\in\{0.5, 0.75, 1, 1.25\}$ (b) and $\tau_v=5$ and $\tau_n\in\{1.2, 1.35, 1.4, 1.6\}$ (d) from simulations (thin lines) and small noise calculation (thick lines).}
\label{fig:theory_no_noise}
\end{figure}

\begin{figure}[t]
\begin{center}
\includegraphics[scale=.9]{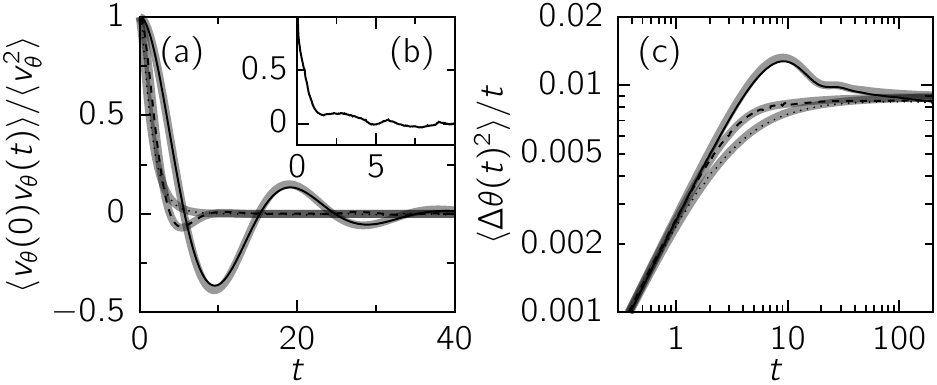}
\end{center}
\vspace{-3mm}
\caption{{\bf Time dependent correlations in the climbing state.} 
(a) Temporal correlations of the azimuthal velocity, and (c) MSD, for $\tau_n=2.5$, $D=0.01$ and $\tau_v=0.2$, $1$ and $5$ (dotted, dashed and solid lines, respectively) in the simulations; thick gray lines show the corresponding weak noise calculation.
(b) Experimental correlations of the azimuthal velocity in the climbing state (Fig.~\ref{fig:exp1}(a-c)).}
\label{fig:stats}
\end{figure}

To rationalize these behaviors, we first study Eqs.~(\ref{eq:motion_v}, \ref{eq:motion_n}) without noise ($D=0$).
We write the equations of motion in polar coordinates: $\dot r=v_r$ and
\begin{align}
\tau_v\left(\dot v_r-\frac{v_\theta^2}{r}\right) & = \cos(\phi)-v_r-r,\label{eq:polar1}\\
\tau_v\left(\dot v_\theta+\frac{v_rv_\theta}{r}\right) & = \sin(\phi)-v_\theta,\\
\tau_n\left(\dot\phi+\frac{v_\theta}{r}\right) & = \cos(\phi)v_\theta-\sin(\phi) v_r,\label{eq:polar3}
\end{align}
We look for stationary solutions, where $v_r=0$, $\dot v_\theta=0$ and $\dot\phi=0$, leading to
\begin{align}
\frac{\tau_v v_\theta^2}{r} & = r-\cos(\phi),\label{eq:stat1}\\
v_\theta & = \sin(\phi),\\
\frac{\tau_n v_\theta}{r} & = \cos(\phi)v_\theta.\label{eq:stat3}
\end{align}
A trivial solution is $r=1$, $v_\theta=0$, $\phi=0$, which corresponds to the climbing state.
When $v_\theta\neq 0$, Eq.~(\ref{eq:stat3}) simplifies and Eqs.~(\ref{eq:stat1}-\ref{eq:stat3}) combine into a closed equation for $\phi$; introducing $u=\cos(\phi)^2$, it reads
\begin{equation}
u^2-\left(1+\frac{\tau_n}{\tau_v} \right)u+\frac{\tau_n^2}{\tau_v}=0.
\end{equation}
This equation has a solution  $u\in[0,1]$, which describes the orbiting state, only if
\begin{equation}
\tau_n\leq\tau_n^*=\left\{\begin{array}{cl}
1 & \text{if }\tau_v\leq 1,\\
\displaystyle{\frac{\tau_v}{2\sqrt{\tau_v}-1}} & \text{if }\tau_v\geq 1.
\end{array}\right.
\end{equation}
Finally, linearizing Eqs.~(\ref{eq:polar1}-\ref{eq:polar3}) around the stationary solutions, we find that the climbing solution is linearly stable for $\tau_n\geq 1$, and the orbiting solution is linearly stable where it exists~\cite{SM}.
The resulting deterministic phase diagram is shown in Fig.~\ref{fig:theory_no_noise}(a).
The transition from climbing to orbiting is mainly controlled by $\tau_n$. For $\tau_v<1$, the transition is continuous. For $\tau_v>1$, it is discontinuous and both solutions coexist in the region  $\tau_v\geq 1$, $1\leq \tau_n\leq \tau_n^*$. 
This is further illustrated  by the evolution of $|v_{\theta}|$ with $\tau_n$, shown in Figs.~\ref{fig:theory_no_noise}(b,d) for $\tau_v=0.2$ and $\tau_v=5$. Here the steady solutions were obtained starting with two different initial conditions: one close to the climbing solution: $\rr_0=(1,0)$, $\vv_0=(1,0)$, and $\phi_0=10^{-4}$ ($IC_1$), the other, $IC_2$, close to the orbiting solution: $\rr_0=(1,0)$, $\vv_0=(0,1)$, and $\phi_0=1.6$.
The evolution of the distribution of azimuthal velocity through the transition (Fig.~\ref{fig:stats}(c,e)) also reflects its continuous or discontinuous character: 
for $\tau_v=0.2$, the most probable value of $v_\theta$ decays continuously from a positive value to zero as $\tau_n$ is increased, whereas it jumps abruptly for $\tau_v=5$. 

It was shown that for large noise, the probability density recovers a Gaussian shape
centered on the center of the parabola~\cite{Solon:2015jd,Basu:2018wo}.
At low noise, the fluctuations $X=(\delta r, v_r, \delta v_\theta, \delta\phi)$ around the stationnary solutions follow $\dot X=-AX+\Xi$ where $A$ is a matrix, whose eigenvalues have a negative real part in a stable state, and $\Xi$ is the Gaussian noise vector associated to $\xi$~\cite{SM}.
Hence, $X(t)$ is Gaussian and can be characterized completely, giving access to the distributions of the position or the azimuthal velocity (Fig.~\ref{fig:numerics}(b,d,h,j), Fig.~\ref{fig:theory_no_noise}(c,e)) and to the azimuthal velocity-orientation correlations (Fig.~\ref{fig:numerics}(e,k))~\cite{SM}.
Notably, in the climbing state, the radius $r$ and radial velocity $v_r$ decouple from the azimuthal velocity $v_\theta$ and orientation $\phi$, which are subject to the angular noise, explaining the very small fluctuations of the radius (Fig.~\ref{fig:numerics}(a,b)).
In the climbing state, we have access to the two regimes of the MSD through the temporal correlations of the azimuthal velocity, $\langle v_\theta(0) v_\theta(t) \rangle$ (Fig.~\ref{fig:numerics}(f)). 
The correlations are proportional to $D$ and the crossover from the sub-diffusive to the diffusive regime does not depend on the noise amplitude. 
Interestingly, for large enough $\tau_n$ and $\tau_v$, the eigenvalues of the matrix $A$ have an imaginary part, leading to oscillations in the velocity correlation (Fig.~\ref{fig:stats}(a)), which translate into a non-trivial ballistic-diffusive crossover of the MSD~(Fig.~\ref{fig:stats}(c)).
We could check that such oscillations are present in the experimental data (see Fig.~\ref{fig:stats}(b)), showing that inertia matters in the experiments.
In the orbiting state, the inversion of the rotation direction is not perturbative and the rate of inversion cannot be computed using this approach. We however capture the ballistic part of the MSD (Fig.~\ref{fig:numerics}(l)).

Altogether using a model of self-propelled particles, previously used to describe active walkers, we have recovered our main experimental observation, namely that an hexbug running in a parabola exhibits two very different dynamical regimes, an orbiting one and a climbing one. 
Furthermore, we understand why the hexbug transits from the orbiting to the climbing regime when the battery gets low: when $F_0$ decreases, $\tau_n$ increases, while $\tau_v$ remains independent of $F_0$, hence the observed transition.
The fact that intermittent dynamics between the two regimes were observed experimentally is also consistent with the existence of the coexistence region for large enough inertia.
Finally the model is able to predict non trivial temporal correlations, the existence of which is confirmed experimentally.

Let us conclude with a few remarks.
First, alternative models including self-alignment have been tested, for instance with solid friction, but none of them reproduces the observed phenomenology~\cite{SM}, showing that our experiment puts strong constraints on the equations describing the motion of the self-propelled particle.
Second, the dynamics described here is in sharp contrast with that of  ABP, for which only climbing dynamics exists. 
This is not related to the absence of inertia in the ABP dynamics: in the limit $\tau_v\to 0$, Eq.~(\ref{eq:motion_v}) becomes $\vv=\nn-\rr$, which is the equation for the velocity of ABP, and the resulting equation for the orientation is
\begin{equation}
\tau_n \dot\nn = -(\nn\times\rr)\times\nn+\sqrt{2D}\xi\nn_\perp.
\end{equation}
The climbing to orbiting transition still exists: the particle rotates for $\tau_n<\tau_n^*=1$ and the azimuthal velocity is $|v_\theta| = \sqrt{1-\tau_n}$. As a consequence, it is truly the self aligning property of $\nn$ towards $\vv$ that is responsible for the presence of the two dynamical regimes.
Finally, as soon as the design of an active particle includes an embedded orientation  $\nn$, as it is the case for Janus particles, a coupling with its velocity, as the one discussed here, could {\it in principle} take place. This is a matter of importance, since it is known that such a coupling is prone to induce collective motion~\cite{Lam:2015bp}. 
The present study offers a way to discriminate between the systems where such self-alignment takes place. 
Once in a harmonic potential a particle with self-alignment should present orbiting solutions.  
From that point of view, it is interesting to remark that the orbiting solution exists, however small the stiffness of the harmonic potential $\kappa$ provided that the rotational diffusion coefficient $D/\tau_n^2 = \alpha\gamma/(\tau^2\kappa)$ remains small enough.
As a matter of fact, a harmonic trap is not the only way to probe the self-alignment of the particle: confining the particle with a hard wall of radius $R_w$ (in dimensionless units), an orbiting solution exists if $\tau_n<R_w$, which slides along the wall at a velocity $v_\parallel=\sqrt{1-(\tau_n/R_w)^2}$~\cite{SM}.
This suggests that the presence of such a coupling could be investigated in a number of active systems, not only those using self-propelled particles similar to the present hexbugs~\cite{Giomi:2012js,Deblais:2018dv}, but also those in the colloidal realm, using for instance acoustic traps~\cite{DeDier:2016he} or simply hard wall circular confinement.



\begin{thebibliography}{20}%
\makeatletter
\providecommand \@ifxundefined [1]{%
 \@ifx{#1\undefined}
}%
\providecommand \@ifnum [1]{%
 \ifnum #1\expandafter \@firstoftwo
 \else \expandafter \@secondoftwo
 \fi
}%
\providecommand \@ifx [1]{%
 \ifx #1\expandafter \@firstoftwo
 \else \expandafter \@secondoftwo
 \fi
}%
\providecommand \natexlab [1]{#1}%
\providecommand \enquote  [1]{``#1''}%
\providecommand \bibnamefont  [1]{#1}%
\providecommand \bibfnamefont [1]{#1}%
\providecommand \citenamefont [1]{#1}%
\providecommand \href@noop [0]{\@secondoftwo}%
\providecommand \href [0]{\begingroup \@sanitize@url \@href}%
\providecommand \@href[1]{\@@startlink{#1}\@@href}%
\providecommand \@@href[1]{\endgroup#1\@@endlink}%
\providecommand \@sanitize@url [0]{\catcode `\\12\catcode `\$12\catcode
  `\&12\catcode `\#12\catcode `\^12\catcode `\_12\catcode `\%12\relax}%
\providecommand \@@startlink[1]{}%
\providecommand \@@endlink[0]{}%
\providecommand \url  [0]{\begingroup\@sanitize@url \@url }%
\providecommand \@url [1]{\endgroup\@href {#1}{\urlprefix }}%
\providecommand \urlprefix  [0]{URL }%
\providecommand \Eprint [0]{\href }%
\providecommand \doibase [0]{http://dx.doi.org/}%
\providecommand \selectlanguage [0]{\@gobble}%
\providecommand \bibinfo  [0]{\@secondoftwo}%
\providecommand \bibfield  [0]{\@secondoftwo}%
\providecommand \translation [1]{[#1]}%
\providecommand \BibitemOpen [0]{}%
\providecommand \bibitemStop [0]{}%
\providecommand \bibitemNoStop [0]{.\EOS\space}%
\providecommand \EOS [0]{\spacefactor3000\relax}%
\providecommand \BibitemShut  [1]{\csname bibitem#1\endcsname}%
\let\auto@bib@innerbib\@empty
\bibitem [{\citenamefont {Bechinger}\ \emph {et~al.}(2016)\citenamefont
  {Bechinger}, \citenamefont {Di~Leonardo}, \citenamefont {L{\"o}wen},
  \citenamefont {Reichhardt}, \citenamefont {Volpe},\ and\ \citenamefont
  {Volpe}}]{Bechinger:2016cf}%
  \BibitemOpen
  \bibfield  {author} {\bibinfo {author} {\bibfnamefont {C.}~\bibnamefont
  {Bechinger}}, \bibinfo {author} {\bibfnamefont {R.}~\bibnamefont
  {Di~Leonardo}}, \bibinfo {author} {\bibfnamefont {H.}~\bibnamefont
  {L{\"o}wen}}, \bibinfo {author} {\bibfnamefont {C.}~\bibnamefont
  {Reichhardt}}, \bibinfo {author} {\bibfnamefont {G.}~\bibnamefont {Volpe}}, \
  and\ \bibinfo {author} {\bibfnamefont {G.}~\bibnamefont {Volpe}},\ }\bibfield
   {title} {\enquote {\bibinfo {title} {{Active Particles in Complex and
  Crowded Environments}},}\ }\href@noop {} {\bibfield  {journal} {\bibinfo
  {journal} {Rev. Mod. Phys.}\ }\textbf {\bibinfo {volume} {88}},\ \bibinfo
  {pages} {1--50} (\bibinfo {year} {2016})}\BibitemShut {NoStop}%
\bibitem [{\citenamefont {Fily}\ \emph {et~al.}(2014)\citenamefont {Fily},
  \citenamefont {Baskaran},\ and\ \citenamefont {Hagan}}]{Fily:2014gy}%
  \BibitemOpen
  \bibfield  {author} {\bibinfo {author} {\bibfnamefont {Y.}~\bibnamefont
  {Fily}}, \bibinfo {author} {\bibfnamefont {A.}~\bibnamefont {Baskaran}}, \
  and\ \bibinfo {author} {\bibfnamefont {M.~F.}\ \bibnamefont {Hagan}},\
  }\bibfield  {title} {\enquote {\bibinfo {title} {{Dynamics of self-propelled
  particles under strong confinement}},}\ }\href@noop {} {\bibfield  {journal}
  {\bibinfo  {journal} {Soft Matter}\ }\textbf {\bibinfo {volume} {10}},\
  \bibinfo {pages} {5609--5617} (\bibinfo {year} {2014})}\BibitemShut {NoStop}%
\bibitem [{\citenamefont {Yan}\ and\ \citenamefont {Brady}(2015)}]{Yan:2015cf}%
  \BibitemOpen
  \bibfield  {author} {\bibinfo {author} {\bibfnamefont {W.}~\bibnamefont
  {Yan}}\ and\ \bibinfo {author} {\bibfnamefont {J.~F.}\ \bibnamefont
  {Brady}},\ }\bibfield  {title} {\enquote {\bibinfo {title} {{The force on a
  boundary in active matter}},}\ }\href@noop {} {\bibfield  {journal} {\bibinfo
   {journal} {J. Fluid Mech.}\ }\textbf {\bibinfo {volume} {785}},\ \bibinfo
  {pages} {R1--11} (\bibinfo {year} {2015})}\BibitemShut {NoStop}%
\bibitem [{\citenamefont {Solon}\ \emph
  {et~al.}(2015{\natexlab{a}})\citenamefont {Solon}, \citenamefont
  {Stenhammar}, \citenamefont {Wittkowski},\ and\ \citenamefont
  {Kardar}}]{Solon:2015hza}%
  \BibitemOpen
  \bibfield  {author} {\bibinfo {author} {\bibfnamefont {A.~P.}\ \bibnamefont
  {Solon}}, \bibinfo {author} {\bibfnamefont {J.}~\bibnamefont {Stenhammar}},
  \bibinfo {author} {\bibfnamefont {R.}~\bibnamefont {Wittkowski}}, \ and\
  \bibinfo {author} {\bibfnamefont {M.}~\bibnamefont {Kardar}},\ }\bibfield
  {title} {\enquote {\bibinfo {title} {{Pressure and phase equilibria in
  interacting active brownian spheres}},}\ }\href@noop {} {\bibfield  {journal}
  {\bibinfo  {journal} {Phys. Rev. Lett.}\ }\textbf {\bibinfo {volume} {114}}
  (\bibinfo {year} {2015}{\natexlab{a}})}\BibitemShut {NoStop}%
\bibitem [{\citenamefont {Nikola}\ \emph {et~al.}(2016)\citenamefont {Nikola},
  \citenamefont {Solon}, \citenamefont {Kafri}, \citenamefont {Kardar},
  \citenamefont {Tailleur},\ and\ \citenamefont {Voituriez}}]{Nikola:2016jca}%
  \BibitemOpen
  \bibfield  {author} {\bibinfo {author} {\bibfnamefont {N.}~\bibnamefont
  {Nikola}}, \bibinfo {author} {\bibfnamefont {A.~P.}\ \bibnamefont {Solon}},
  \bibinfo {author} {\bibfnamefont {Y.}~\bibnamefont {Kafri}}, \bibinfo
  {author} {\bibfnamefont {M.}~\bibnamefont {Kardar}}, \bibinfo {author}
  {\bibfnamefont {J.}~\bibnamefont {Tailleur}}, \ and\ \bibinfo {author}
  {\bibfnamefont {R.}~\bibnamefont {Voituriez}},\ }\bibfield  {title} {\enquote
  {\bibinfo {title} {{Active Particles with Soft and Curved Walls: Equation of
  State, Ratchets, and Instabilities}},}\ }\href@noop {} {\bibfield  {journal}
  {\bibinfo  {journal} {Phys. Rev. Lett.}\ }\textbf {\bibinfo {volume} {117}},\
  \bibinfo {pages} {098001--5} (\bibinfo {year} {2016})}\BibitemShut {NoStop}%
\bibitem [{\citenamefont {De~Dier}\ \emph {et~al.}(2016)\citenamefont
  {De~Dier}, \citenamefont {Vermant}, \citenamefont {Brady},\ and\
  \citenamefont {Takatori}}]{DeDier:2016he}%
  \BibitemOpen
  \bibfield  {author} {\bibinfo {author} {\bibfnamefont {R.}~\bibnamefont
  {De~Dier}}, \bibinfo {author} {\bibfnamefont {J.}~\bibnamefont {Vermant}},
  \bibinfo {author} {\bibfnamefont {J.~F.}\ \bibnamefont {Brady}}, \ and\
  \bibinfo {author} {\bibfnamefont {S.~C.}\ \bibnamefont {Takatori}},\
  }\bibfield  {title} {\enquote {\bibinfo {title} {{Acoustic trapping of active
  matter}},}\ }\href@noop {} {\bibfield  {journal} {\bibinfo  {journal} {N.
  Commun.}\ }\textbf {\bibinfo {volume} {7}},\ \bibinfo {pages} {1--7}
  (\bibinfo {year} {2016})}\BibitemShut {NoStop}%
\bibitem [{\citenamefont {Junot}\ \emph {et~al.}(2017)\citenamefont {Junot},
  \citenamefont {Briand}, \citenamefont {Ledesma-Alonso},\ and\ \citenamefont
  {Dauchot}}]{Junot:2017fd}%
  \BibitemOpen
  \bibfield  {author} {\bibinfo {author} {\bibfnamefont {G.}~\bibnamefont
  {Junot}}, \bibinfo {author} {\bibfnamefont {G.}~\bibnamefont {Briand}},
  \bibinfo {author} {\bibfnamefont {R.}~\bibnamefont {Ledesma-Alonso}}, \ and\
  \bibinfo {author} {\bibfnamefont {O.}~\bibnamefont {Dauchot}},\ }\bibfield
  {title} {\enquote {\bibinfo {title} {{Active versus Passive Hard Disks
  against a Membrane: Mechanical Pressure and Instability}},}\ }\href@noop {}
  {\bibfield  {journal} {\bibinfo  {journal} {Phys. Rev. Lett.}\ } (\bibinfo
  {year} {2017})}\BibitemShut {NoStop}%
\bibitem [{\citenamefont {Deblais}\ \emph {et~al.}(2018)\citenamefont
  {Deblais}, \citenamefont {Barois}, \citenamefont {Guérin}, \citenamefont
  {Delville}, \citenamefont {Vaudaine}, \citenamefont {Lintuvuori},
  \citenamefont {Boudet}, \citenamefont {Baret},\ and\ \citenamefont
  {Kellay}}]{Deblais:2018dv}%
  \BibitemOpen
  \bibfield  {author} {\bibinfo {author} {\bibfnamefont {A.}~\bibnamefont
  {Deblais}}, \bibinfo {author} {\bibfnamefont {T.}~\bibnamefont {Barois}},
  \bibinfo {author} {\bibfnamefont {T.}~\bibnamefont {Guérin}}, \bibinfo
  {author} {\bibfnamefont {P.~H.}\ \bibnamefont {Delville}}, \bibinfo {author}
  {\bibfnamefont {R.}~\bibnamefont {Vaudaine}}, \bibinfo {author}
  {\bibfnamefont {J.~S.}\ \bibnamefont {Lintuvuori}}, \bibinfo {author}
  {\bibfnamefont {J.~F.}\ \bibnamefont {Boudet}}, \bibinfo {author}
  {\bibfnamefont {J.~C.}\ \bibnamefont {Baret}}, \ and\ \bibinfo {author}
  {\bibfnamefont {H.}~\bibnamefont {Kellay}},\ }\bibfield  {title} {\enquote
  {\bibinfo {title} {{Boundaries Control Collective Dynamics of Inertial
  Self-Propelled Robots}},}\ }\href@noop {} {\bibfield  {journal} {\bibinfo
  {journal} {Phys. Rev. Lett.}\ }\textbf {\bibinfo {volume} {120}},\ \bibinfo
  {pages} {188002} (\bibinfo {year} {2018})}\BibitemShut {NoStop}%
\bibitem [{\citenamefont {Solon}\ \emph
  {et~al.}(2015{\natexlab{b}})\citenamefont {Solon}, \citenamefont {Cates},\
  and\ \citenamefont {Tailleur}}]{Solon:2015jd}%
  \BibitemOpen
  \bibfield  {author} {\bibinfo {author} {\bibfnamefont {A.~P.}\ \bibnamefont
  {Solon}}, \bibinfo {author} {\bibfnamefont {M.~E.}\ \bibnamefont {Cates}}, \
  and\ \bibinfo {author} {\bibfnamefont {J.}~\bibnamefont {Tailleur}},\
  }\bibfield  {title} {\enquote {\bibinfo {title} {{Active brownian particles
  and run-and-tumble particles: A comparative study}},}\ }\href@noop {}
  {\bibfield  {journal} {\bibinfo  {journal} {Eur. Phys. J. ST}\ }\textbf
  {\bibinfo {volume} {224}},\ \bibinfo {pages} {1231--1262} (\bibinfo {year}
  {2015}{\natexlab{b}})}\BibitemShut {NoStop}%
\bibitem [{\citenamefont {Fodor}\ and\ \citenamefont
  {Marchetti}(2018)}]{Fodor:2018en}%
  \BibitemOpen
  \bibfield  {author} {\bibinfo {author} {\bibfnamefont {É.}\ \bibnamefont
  {Fodor}}\ and\ \bibinfo {author} {\bibfnamefont {M.~C.}\ \bibnamefont
  {Marchetti}},\ }\bibfield  {title} {\enquote {\bibinfo {title} {{The
  statistical physics of active matter: From self-catalytic colloids to living
  cells}},}\ }\href@noop {} {\bibfield  {journal} {\bibinfo  {journal} {Physica
  A}\ ,\ \bibinfo {pages} {1--15}} (\bibinfo {year} {2018})}\BibitemShut
  {NoStop}%
\bibitem [{\citenamefont {Maggi}\ \emph {et~al.}(2015)\citenamefont {Maggi},
  \citenamefont {Marconi}, \citenamefont {Gnan},\ and\ \citenamefont
  {Di~Leonardo}}]{Maggi:2015jl}%
  \BibitemOpen
  \bibfield  {author} {\bibinfo {author} {\bibfnamefont {C.}~\bibnamefont
  {Maggi}}, \bibinfo {author} {\bibfnamefont {U.~M.~B.}\ \bibnamefont
  {Marconi}}, \bibinfo {author} {\bibfnamefont {N.}~\bibnamefont {Gnan}}, \
  and\ \bibinfo {author} {\bibfnamefont {R.}~\bibnamefont {Di~Leonardo}},\
  }\bibfield  {title} {\enquote {\bibinfo {title} {{Multidimensional stationary
  probability distribution for interacting active particles}},}\ }\href@noop {}
  {\bibfield  {journal} {\bibinfo  {journal} {Sci. Rep.}\ ,\ \bibinfo {pages}
  {1--7}} (\bibinfo {year} {2015})}\BibitemShut {NoStop}%
\bibitem [{\citenamefont {Basu}\ \emph {et~al.}(2018)\citenamefont {Basu},
  \citenamefont {Majumdar}, \citenamefont {Rosso},\ and\ \citenamefont
  {Schehr}}]{Basu:2018wo}%
  \BibitemOpen
  \bibfield  {author} {\bibinfo {author} {\bibfnamefont {U.}~\bibnamefont
  {Basu}}, \bibinfo {author} {\bibfnamefont {S.~N.}\ \bibnamefont {Majumdar}},
  \bibinfo {author} {\bibfnamefont {A.}~\bibnamefont {Rosso}}, \ and\ \bibinfo
  {author} {\bibfnamefont {G.}~\bibnamefont {Schehr}},\ }\bibfield  {title}
  {\enquote {\bibinfo {title} {{Active Brownian Motion in Two Dimensions}},}\
  }\href@noop {} {\bibfield  {journal} {\bibinfo  {journal} {arXiv}\ }
  (\bibinfo {year} {2018})},\ \Eprint {http://arxiv.org/abs/1804.09027v1}
  {1804.09027v1} \BibitemShut {NoStop}%
\bibitem [{hex()}]{hexbug}%
  \BibitemOpen
  \bibfield  {title} {\enquote {\bibinfo {title} {{Hexbug is a toy automat
  brand developed and distributed by Innovation First
  https://www.hexbug.com/}},}\ }\href@noop {} {\ }\BibitemShut {NoStop}%
\bibitem [{\citenamefont {Scholz}\ \emph {et~al.}(2018)\citenamefont {Scholz},
  \citenamefont {Jahanshahi}, \citenamefont {Ldov},\ and\ \citenamefont
  {L{\"{o}}wen}}]{Scholz2018}%
  \BibitemOpen
  \bibfield  {author} {\bibinfo {author} {\bibfnamefont {C.}~\bibnamefont
  {Scholz}}, \bibinfo {author} {\bibfnamefont {S.}~\bibnamefont {Jahanshahi}},
  \bibinfo {author} {\bibfnamefont {A.}~\bibnamefont {Ldov}}, \ and\ \bibinfo
  {author} {\bibfnamefont {H.}~\bibnamefont {L{\"{o}}wen}},\ }\bibfield
  {title} {\enquote {\bibinfo {title} {{Inertial delay of self-propelled
  particles}},}\ }\href@noop {} {\bibfield  {journal} {\bibinfo  {journal}
  {{ArXiv}}\ } (\bibinfo {year} {2018})},\ \Eprint
  {http://arxiv.org/abs/{1807.04357}} {{1807.04357}} \BibitemShut {NoStop}%
\bibitem [{\citenamefont {Weber}\ \emph {et~al.}(2013)\citenamefont {Weber},
  \citenamefont {Hanke}, \citenamefont {Deseigne}, \citenamefont {L{\'e}onard},
  \citenamefont {Dauchot}, \citenamefont {Frey},\ and\ \citenamefont
  {Chat{\'e}}}]{Weber:2013bj}%
  \BibitemOpen
  \bibfield  {author} {\bibinfo {author} {\bibfnamefont {C.~A.}\ \bibnamefont
  {Weber}}, \bibinfo {author} {\bibfnamefont {T.}~\bibnamefont {Hanke}},
  \bibinfo {author} {\bibfnamefont {J.}~\bibnamefont {Deseigne}}, \bibinfo
  {author} {\bibfnamefont {S.}~\bibnamefont {L{\'e}onard}}, \bibinfo {author}
  {\bibfnamefont {O.}~\bibnamefont {Dauchot}}, \bibinfo {author} {\bibfnamefont
  {E.}~\bibnamefont {Frey}}, \ and\ \bibinfo {author} {\bibfnamefont
  {H.}~\bibnamefont {Chat{\'e}}},\ }\bibfield  {title} {\enquote {\bibinfo
  {title} {{Long-Range Ordering of Vibrated Polar Disks}},}\ }\href@noop {}
  {\bibfield  {journal} {\bibinfo  {journal} {Phys. Rev. Lett.}\ }\textbf
  {\bibinfo {volume} {110}},\ \bibinfo {pages} {208001} (\bibinfo {year}
  {2013})}\BibitemShut {NoStop}%
\bibitem [{\citenamefont {Bore}\ \emph {et~al.}(2016)\citenamefont {Bore},
  \citenamefont {Schindler}, \citenamefont {Nguyen Thu~Lam}, \citenamefont
  {Bertin},\ and\ \citenamefont {Dauchot}}]{LolandBore:2016if}%
  \BibitemOpen
  \bibfield  {author} {\bibinfo {author} {\bibfnamefont {S.~L.}\ \bibnamefont
  {Bore}}, \bibinfo {author} {\bibfnamefont {M.}~\bibnamefont {Schindler}},
  \bibinfo {author} {\bibfnamefont {K.-D.}\ \bibnamefont {Nguyen Thu~Lam}},
  \bibinfo {author} {\bibfnamefont {É.~M.}\ \bibnamefont {Bertin}}, \ and\
  \bibinfo {author} {\bibfnamefont {O.}~\bibnamefont {Dauchot}},\ }\bibfield
  {title} {\enquote {\bibinfo {title} {{Coupling spin to velocity: collective
  motion of Hamiltonian polar particles}},}\ }\href@noop {} {\bibfield
  {journal} {\bibinfo  {journal} {J. Stat. Mech.}\ }\textbf {\bibinfo {volume}
  {2016}},\ \bibinfo {pages} {033305} (\bibinfo {year} {2016})}\BibitemShut
  {NoStop}%
\bibitem [{\citenamefont {Nguyen Thu~Lam}\ \emph
  {et~al.}(2015{\natexlab{a}})\citenamefont {Nguyen Thu~Lam}, \citenamefont
  {Schindler},\ and\ \citenamefont {Dauchot}}]{NguyenThuLam2015}%
  \BibitemOpen
  \bibfield  {author} {\bibinfo {author} {\bibfnamefont {K.-D.}\ \bibnamefont
  {Nguyen Thu~Lam}}, \bibinfo {author} {\bibfnamefont {M.}~\bibnamefont
  {Schindler}}, \ and\ \bibinfo {author} {\bibfnamefont {O.}~\bibnamefont
  {Dauchot}},\ }\bibfield  {title} {\enquote {\bibinfo {title} {{Self-propelled
  hard disks: implicit alignment and transition to collective motion}},}\
  }\href@noop {} {\bibfield  {journal} {\bibinfo  {journal} {{New J. Phys.}}\
  }\textbf {\bibinfo {volume} {17}},\ \bibinfo {pages} {113056} (\bibinfo
  {year} {2015}{\natexlab{a}})}\BibitemShut {NoStop}%
\bibitem [{SM()}]{SM}%
  \BibitemOpen
  \bibfield  {title} {\enquote {\bibinfo {title} {{See supplemental material at
  [url will be inserted by publisher] for detailed calculations.}}}\
  }\href@noop {} {\ }\BibitemShut {NoStop}%
\bibitem [{\citenamefont {Nguyen Thu~Lam}\ \emph
  {et~al.}(2015{\natexlab{b}})\citenamefont {Nguyen Thu~Lam}, \citenamefont
  {Schindler},\ and\ \citenamefont {Dauchot}}]{Lam:2015bp}%
  \BibitemOpen
  \bibfield  {author} {\bibinfo {author} {\bibfnamefont {K.-D.}\ \bibnamefont
  {Nguyen Thu~Lam}}, \bibinfo {author} {\bibfnamefont {M.}~\bibnamefont
  {Schindler}}, \ and\ \bibinfo {author} {\bibfnamefont {O.}~\bibnamefont
  {Dauchot}},\ }\bibfield  {title} {\enquote {\bibinfo {title} {{Self-propelled
  hard disks: implicit alignment and transition to collective motion}},}\
  }\href@noop {} {\bibfield  {journal} {\bibinfo  {journal} {New J. Phys.}\
  }\textbf {\bibinfo {volume} {17}},\ \bibinfo {pages} {113056} (\bibinfo
  {year} {2015}{\natexlab{b}})}\BibitemShut {NoStop}%
\bibitem [{\citenamefont {Giomi}\ \emph {et~al.}(2012)\citenamefont {Giomi},
  \citenamefont {Hawley-Weld},\ and\ \citenamefont {Mahadevan}}]{Giomi:2012js}%
  \BibitemOpen
  \bibfield  {author} {\bibinfo {author} {\bibfnamefont {L.}~\bibnamefont
  {Giomi}}, \bibinfo {author} {\bibfnamefont {N.}~\bibnamefont {Hawley-Weld}},
  \ and\ \bibinfo {author} {\bibfnamefont {L.}~\bibnamefont {Mahadevan}},\
  }\bibfield  {title} {\enquote {\bibinfo {title} {{Swarming, swirling and
  stasis in sequestered bristle-bots}},}\ }\href@noop {} {\bibfield  {journal}
  {\bibinfo  {journal} {Proc. Royal Soc. A}\ }\textbf {\bibinfo {volume}
  {469}},\ \bibinfo {pages} {20120637--20120637} (\bibinfo {year}
  {2012})}\BibitemShut {NoStop}%
\end{thebibliography}
%

\newpage

\section{Stability analysis}\label{sec:stab_analysis}

\subsection{Immobile solution}\label{sec:stab_analysis_immo}

We linearize Eqs.~(\ref{eq:polar1}-\ref{eq:polar3}) around $\vv=0$, $\phi=0$, $r=1$.
At first order in $r_1=r-1$, $v_r$, $v_\theta$, and $\phi$, we get
\begin{align}
\dot r_1 & = v_r,\\
\dot v_r & = -\tau_v^{-1}(r_1+v_r),\\
\dot v_\theta & = \tau_v^{-1}(\phi-v_\theta),\\
\dot \phi & = (\tau_n^{-1}-1)v_\theta.
\end{align}
We introduce the vector
\begin{equation}\label{eq:vector_pert}
X = \begin{pmatrix}
r_1 \\ v_r\\v_{\theta,1}\\\phi_1
\end{pmatrix},
\end{equation}
which evolves according to $\dot X = AX$, where
\begin{equation}\label{eq:lin_stab_climb_diag}
A = \begin{pmatrix}
A_r & 0 \\ 0 & A_\theta
\end{pmatrix}
\end{equation}
with
\begin{align}
A_r & = \begin{pmatrix}
0 & 1\\-\tau_v^{-1}&-\tau_v^{-1}
\end{pmatrix}\\
A_\theta & = \begin{pmatrix}
0 & \tau_n^{-1}-1\\\tau_v^{-1}&-\tau_v^{-1}
\end{pmatrix}.
\end{align}

Equation (\ref{eq:lin_stab_climb_diag}) means that the radial and azimuthal directions decouple and can be analyzed independently.
In the radial direction, the eigenvalues of $A_r$ are
\begin{equation}
\mathrm{Spec}(A_r)=\left\{-\frac{\tau_v^{-1}\pm\sqrt{\tau_v^{-2}-4\tau_v^{-1}}}{2} \right\},
\end{equation}
and their real part is negative.
In the azimuthal direction, the eigenvalues of $A_\theta$ are
\begin{equation}
\mathrm{Spec}(A_\theta)=\left\{-\frac{\tau_v^{-1}\pm\sqrt{\tau_v^{-2}+4(\tau_n^{-1}-1)\tau_v^{-1}}}{2} \right\},
\end{equation}
The real part of these eigenvalues is negative if $\tau_n^{-1}-1<0$, which corresponds to
\begin{equation}
\tau_n>1.
\end{equation}
This is the condition for the immobile solutions to be stable.

\begin{figure}
\begin{center}
\includegraphics[scale=.9]{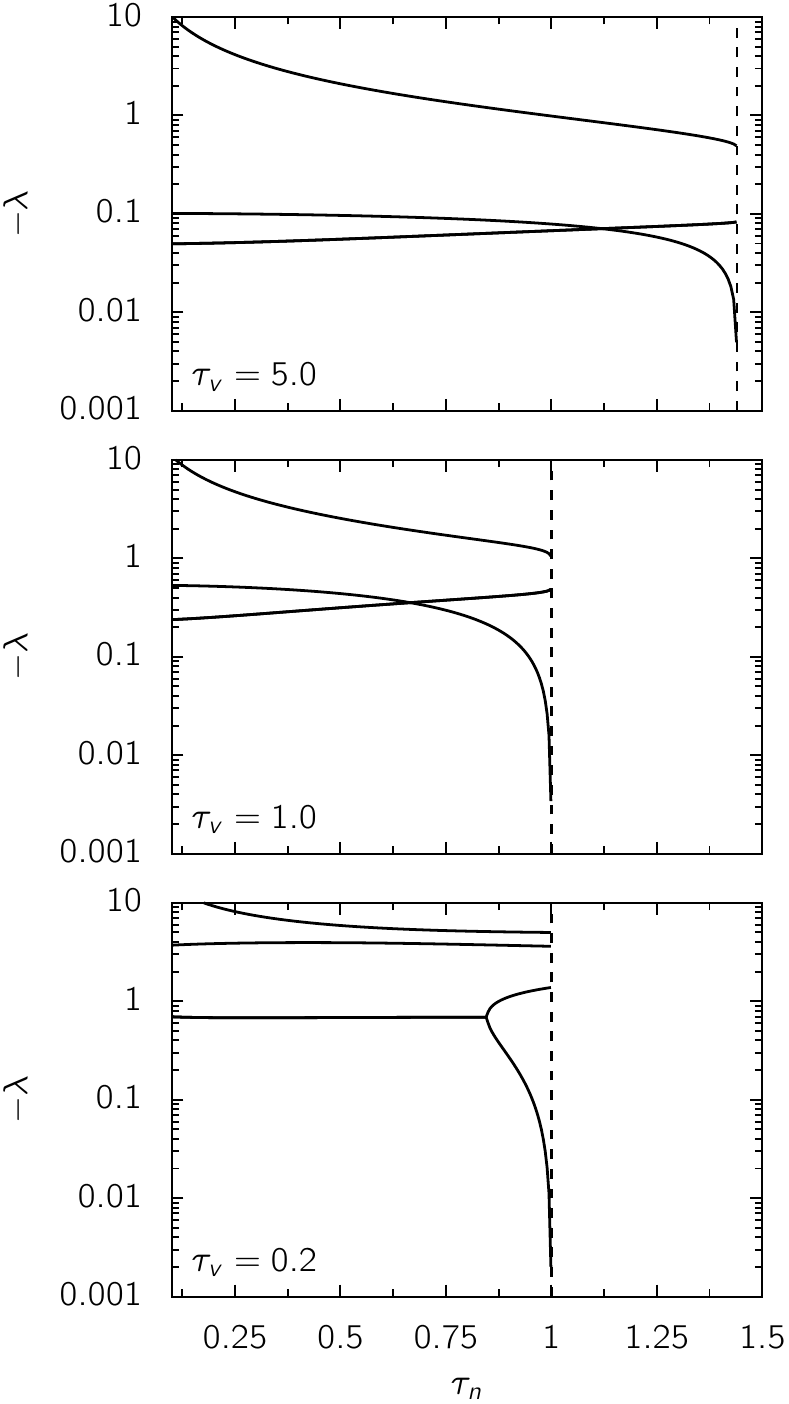}
\end{center}
\caption{Minus the real part of the eigenvalues of the linear response matrix (Eq.~(\ref{eq:lin_matrix_rot})) around the orbiting solution as a function of $\tau_n$ for different values of $\tau_v$. 
The dashed line indicates the critical value of $\tau_n$ above which no orbiting solution exists.}
\label{fig:stab_rot}
\end{figure}

\begin{widetext}

\subsection{Orbiting solution}\label{sec:stab_analysis_rot}

We write $r=r_0+r_1$, $v_\theta=v_{\theta,0}+v_{\theta,1}$ and $\phi=\phi_0+\phi_1$, where the quantities of order 0 satisfy the relations (\ref{eq:stat1}-\ref{eq:stat3}) with $\omega=v_{\theta,0}/r_0\neq 0$.
We expand Eqs.~(\ref{eq:polar1}-\ref{eq:polar3}) at order 1 in the quantities $r_1$, $v_r$, $v_{\theta,1}$, $\phi_1$~:
\begin{align}
\tau_v \left(\dot v_r-\frac{2v_{\theta,0}}{r_0}v_{\theta,1}+\frac{v_{\theta,0}^2}{r_0^2}r_1 \right) & = -r_1-v_r-\sin(\phi_0)\phi_1,\\
\tau_v \left(\dot v_{\theta,1}+\frac{v_{\theta,0}}{r_0}v_r \right) & = -v_{\theta,1}+\cos(\phi_0)\phi_1,\\
\tau_n \left(\dot\phi_1+\frac{v_{\theta,1}}{r_0}-\frac{v_{\theta,0}}{r_0^2}r_1 \right) & = -\sin(\phi_0)v_r+\cos(\phi_0)v_{\theta,1}-\sin(\phi_0)v_{\theta,0}\phi_1.
\end{align}
The last equation is $\dot r_1=v_r$.
Here the matrix $A$ is
\begin{equation}
A = \begin{pmatrix}
0 & 1 & 0 & 0\\
-\tau_v^{-1}-\frac{v_{\theta,0}^2}{r_0^2} & -\tau_v^{-1} & \frac{2v_{\theta,0}}{r_0} & -\tau_v^{-1}\sin(\phi_0)\\
0 & -\frac{v_{\theta,0}}{r_0} & -\tau_v^{-1} & \tau_v^{-1}\cos(\phi_0)\\
\frac{v_{\theta,0}}{r_0^2} & -\tau_n^{-1}\sin(\phi_0) & \tau_n^{-1}\cos(\phi_0)-\frac{1}{r_0} & -\tau_n^{-1}\sin(\phi_0)v_{\theta,0}
\end{pmatrix}.
\end{equation}
This matrix can be written in a more compact form using $\omega_0=v_{\theta,0}/r_0$, $\sin(\phi_0)=v_{\theta,0}$, $\cos(\phi_0)=\tau_n/r$:
\begin{equation}\label{eq:lin_matrix_rot}
A = \begin{pmatrix}
0 & 1 & 0 & 0\\
-\tau_v^{-1}-\omega_0^2 & -\tau_v^{-1} & 2\omega_0 & -\tau_v^{-1}v_{\theta,0}\\
0 & -\omega_0 & -\tau_v^{-1} & \frac{\tau_n}{\tau_v r_0}\\
\frac{\omega_0}{r_0} & -\tau_n^{-1}v_{\theta,0} & 0 & -\tau_n^{-1}v_{\theta,0}^2
\end{pmatrix}.
\end{equation}
The eigenvalues of this matrix can be computed numerically for given values of $\tau_v$ and $\tau_n$, their real part are plotted in Fig.~\ref{fig:stab_rot}.
It appears that the eigenvalues have a negative real part, with an eigenvalue going to 0 as $\tau_n$ approaches the critical value $\tau_n^*$.

\end{widetext}

\section{Effect of a weak angular noise}\label{}

\subsection{Correlations}\label{}

In presence of noise Eq.~(\ref{eq:polar3}) reads
\begin{equation}\label{eq:polar3_noise}
\tau_n(\dot\phi+\omega) = n_rv_\theta-n_\theta v_r+\xi.
\end{equation}
If the noise is weak, its effect can be obtained from the linear response matrices computed for the stability analysis.
The small deviations (Eq.~(\ref{eq:vector_pert})) now follow
\begin{equation}
\dot X = AX+\Xi,
\end{equation}
where $A$ is the matrix computed in the stability analysis and 
\begin{equation}
\Xi(t)=\tau_n^{-1}\xi(t)\begin{pmatrix}
0\\0\\0\\1
\end{pmatrix}.
\end{equation}
We define the correlations
\begin{equation}
C(\tau)=\langle X(t+\tau)X(t)\transp \rangle.
\end{equation}
Writing the correlation of the noise as
\begin{equation}
\left\langle \Xi(t)\Xi(t')\transp \right\rangle = R\delta(t-t'),
\end{equation}
it is straightforward to show that
\begin{equation}\label{eq:correl_int}
C(\tau) = \int_0^\infty e^{(u+\tau)A}Re^{uA\transp}du.
\end{equation}
This correlation can be computed by diagonalizing $A$:
\begin{equation}
A = PDP^{-1} = \sum_i \lambda_i PE_iP^{-1},
\end{equation}
where $D$ is a diagonal matrix and $E_i$ is a matrix with elements $(E_i)_{kl}=\delta_{ik}\delta_{il}$; $\lambda_i$ are the eigenvalues and $P$ is the change of basis.
Inserting in Eq.~(\ref{eq:correl_int}) and using that the real part of the eigenvalues are negative, we get
\begin{equation}
C(\tau) = -\sum_{i,j} \frac{e^{\tau \lambda_i}}{\lambda_i+\lambda_j} PE_iP^{-1}R{P^{-1}}\transp E_j P\transp.
\end{equation}
Thus all the correlations can be computed.

For the immobile solution, the radial coordinates are not coupled to the orthoradial coordinates and are thus independent of the noise: they do not fluctuate.

\subsection{Mean square angular displacement in the immobile solution}\label{}

The radius is $r=1$ and it does not fluctuate, hence the angular displacement is given by $\dot\theta(t)=v_\theta(t)$.
The MSD is given by
\begin{equation}
\langle \Delta\theta(t)^2 \rangle = 2\int_0^t (t-t')C_{v_\theta}(t') dt'.
\end{equation}
From the calculation above, it follows that the correlation is a sum of exponentials:
\begin{equation}
C_{v_\theta}(t) = \sum_i a_ie^{-b_it},
\end{equation}
leading to
\begin{equation}
\langle \Delta\theta(t)^2 \rangle = 2\sum_i \frac{a_i}{b_i^2} \left(b_i t-1+e^{-b_it} \right),
\end{equation}
which can be evaluated numerically.

\section{Alternative models}\label{}

\subsection{Solid friction}\label{}

The fluid friction of Eq.~(\ref{eq:motion_v}) can be replaced by solid friction:
\begin{equation}
\tau_v\dot\vv = \nn -\hat \vv-\rr.
\end{equation}
In this case the equations for a stationnary orbiting state read:
\begin{align}
\tau_v\omega^2r & = r-\cos(\phi),\\ 
1 &=\sin(\phi),\\
\tau_n\omega & =\omega r\cos(\phi). 
\end{align}
Hence $\phi$ is constrained to $\phi=\pi/2$. 
Without confinement, a solid friction is not compatible with a self-propulsion force: for a stationnary velocity to exist, the friction should increase with the velocity.

\subsection{Alignment with velocity orientation}\label{}

The torque on the orientation is proportionnal to the magnitude of the velocity $\vv$ in Eq.~(\ref{eq:motion_n}).
Alternatively, the torque could depend on the orientation of the velocity only; the equation for the orientation $\nn$ would be replaced by
\begin{equation}
\tau_n\dot\nn = (\nn\times\hat\vv)\times \nn.
\end{equation}
A stationnary orbiting state is now described by
\begin{align}
\tau_v\omega^2r & = r-\cos(\phi),\\
\omega r&=\sin(\phi), \label{eq:hatv_vt}\\
\tau_n\omega & =\cos(\phi). \label{eq:hatv_n}
\end{align}
Combining Eqs.~(\ref{eq:hatv_vt}, \ref{eq:hatv_n}), we get
\begin{equation}
r = \tau_n\tan(\phi).
\end{equation}
In this case the radius should decay to zero as the climbing solution is approached, which is not compatible with the experiments.

\section{Confinement with a curved hard wall}\label{}

If the harmonic confinement term $-\mu \rr$ in Eq.~(\ref{eq:motion_v}) is replaced with a hard wall located on the circle $r=R_w$, the radius in a orbiting state is set to $R_w$, and Eqs.~(\ref{eq:stat1}-\ref{eq:stat3}) are replaced by
\begin{align}
v_\theta & = \sin(\phi),\\
\frac{\tau_n v_\theta}{R_w} & = \cos(\phi)v_\theta. \label{eq:stat_hw_2}
\end{align}
If $v_\theta\neq 0$, $v_\theta$ simplifies in Eq.~(\ref{eq:stat_hw_2}) and there is a solution only if
\begin{equation}
\tau_n<R_w;
\end{equation}
in this case the azimuthal velocity is given by
\begin{equation}
v_\theta=\sqrt{1-\frac{\tau_n^2}{R_w^2}}.
\end{equation}

\end{document}